\documentclass{aa}
\usepackage{graphicx}

\begin{document}
%
%
\title{Mass segregation of different populations inside the cluster NGC6101
\thanks{Based on observations collected 
at the European Southern Observatory, La Silla Chile and on HST observations.
Tables with the x, y coordinates, V magnitudes and (V-I), (B-V) colors 
(for both ground and HST-data), are only available in electronic form 
at the CDS via anonymous ftp.}
}

\author{Marconi G. \inst{1}, Andreuzzi G. \inst{2}, Pulone L.
\inst{2}, Cassisi S. \inst{3}, Testa V. \inst{2}, Buonanno R. \inst{4,2}}

\offprints{G. Marconi}
\institute{European Southern Observatory, Casilla 19001, Santiago, Chile,
email: gmarconi@eso.org\\ 
\and Osservatorio Astronomico di Roma, via Frascati 33, 00040 Monteporzio 
Catone, ITALY\\
\and  Osservatorio Astronomico di Collurania, Via M. Maggini, 64100 Teramo, 
ITALY\\
\and Dipartimento di Fisica, Universit\`a di Roma ``Tor Vergata'', 00173, Roma,
ITALY}

\date{Received / accepted}

\abstract{
We have used ESO telescopes at La Silla and the Hubble Space Telescope 
(HST) in order to obtain accurate B,V,I CCD photometry for the stars 
located within 200$\arcsec$ ($\simeq$ 2 half-mass radii,
$r{\rm _h}$ = 1.71$\arcmin$) from the center of the cluster NGC\,6101.
Color-Magnitude Diagrams (CMDs) extending from the red-giant tip to  
about 5 magnitudes below the main-sequence turnoff MSTO (V = 20.05 $\pm$ 0.05)
have been constructed. 
The following results have been obtained from the analysis of the CMDs:
a) The overall morphology  of the main branches confirms previous results 
from the literature, in particular the existence of a sizeable 
population of 73 ``blue stragglers'' (BSS),  which had been already
partly detected (27).
They are considerably more concentrated than either the subgiant branch (SGB)
or the main sequence (MS) stars, and have the same spatial distribution as the 
horizontal branch (HB) stars (84 $\%$ probability from K-S test). 
An hypothesis on the possible BSS progeny  is also presented.
b)  The HB is narrow and the bulk of stars is blue, as expected for a 
typical metal-poor globular cluster.
c)  The derived magnitudes for the HB and the MSTO, $V_{\rm {ZAHB}}$ = 16.59 
$\pm$ 0.10, $V_{\rm {TO}}$ = 20.05 $\pm$ 0.05, coupled with 
the values E(B-V) = 0.1, [Fe/H] = -1.80, Y = 0.23 yield a distance modulus 
$(m-M)_{\rm V}$ = 16.23 and an age similar to other ``old'' metal-poor globular 
clusters.
In particular, from the comparison with theoretical isochrones, we derive for 
this cluster an age of 13 Gyrs.
d) By using the large statistical sample of Red Giant Branch (RGB) stars, 
we detected with high accuracy the position of the bump in the RGB
luminosity function. This observational feature
has been compared with theoretical prescriptions, yielding a good
agreement within the current theoretical and observational uncertainties.
}

\authorrunning{G. Marconi et al.}
\titlerunning{Mass segregation in NGC 6101}

\maketitle

\section{Introduction}

NGC\,6101 is a ``typical'' metal-poor galactic globular cluster located 
at $\alpha = 16^{\rm h} 20.0^{\rm m}$, $\delta$ = -$72^{{\rm \circ}}$ 
$05^{{\rm \prime}}$, (l = 
$318^{{\rm \circ}}$, b = -$16^{{\rm \circ}}$) with a low central concentration
($log \rho_{\rm 0}$ = 1.57, Djorgovski \cite{djor93}). 

A photographic study of the cluster was done by Alcaino (\cite{alca74}); 
there are only two important photometric studies based on CCD data: 
Sarajedini  $\&$ Da Costa (\cite{sara91}, SDC) and Rosenberg et al. 
(\cite{rose00}).
The most complete analysis was performed by SDC by using B and 
V observations of 4 fields in the external regions of the cluster. 
The main results of SDC are:
(1) NGC\,6101 is a ``normal'' metal-poor globular cluster of an age 
similar to other galactic globular clusters with similar metal abundances;
(2) it contains a conspicuous fraction of BSS (27) with properties 
similar to the BSS of other globular clusters; in particular, from the analysis
of radial distributions, they found that the BSS are more concentrated than 
the SGB stars.

We have obtained new ground-based and HST B,V, I high-quality photometry of 
400 arcsec$^{\rm 2}$ centered on the cluster. The new data-set made it possible to 
re-analyse the stellar content inside $\simeq$ 2 $r_{\rm h}$.
For the sake of comparison, both CCDs studies quoted above have been used.

In  Sect. 2 we describe the observations and the procedures applied 
in order to reduce and calibrate the data. 
Section 3 shows the (V, B-V) and (V, V-I) CMDs. 
The luminosity functions (LFs) and the radial distributions of the stars are 
discussed in  Sects. 4 and 5.
In  Sect. 6 age, distance modulus and metallicity are derived from the
comparison with  theoretical models.
The results are summarized in  Sect. 7.

\section{Observations and data reduction}

\subsection{Observations}
\label{sub:obs}

The ground-based B,V,I images have been obtained at 
ESO/La Silla with the 1.54 Danish telescope and the Direct CCD camera.
A large number of CCD frames of a field centered on NGC\,6101 were obtained 
under good seeing conditions,  during the observing run of July 1995.
The CCD (a Tektronik 1024 with scale 0.37$\arcsec$/pixel and field of view of
400 arcsec$^{\rm 2}$)  sampled very well the seeing profile which 
varies from 0.75$\arcsec$ to 0.95$\arcsec$.
Standard stars for calibration were taken  every night from Landolt's 
regions (Landolt \cite{land92}). 

In order to carefully check the crowding effect in the most central region, 
deep exposures with HST + WFPC2 (with the filters F555W and F814W) 
were obtained for the center of the cluster within the GO 6625 program 
(Cycle 6).
 
\begin{figure}
 \resizebox{\hsize}{!}{\includegraphics{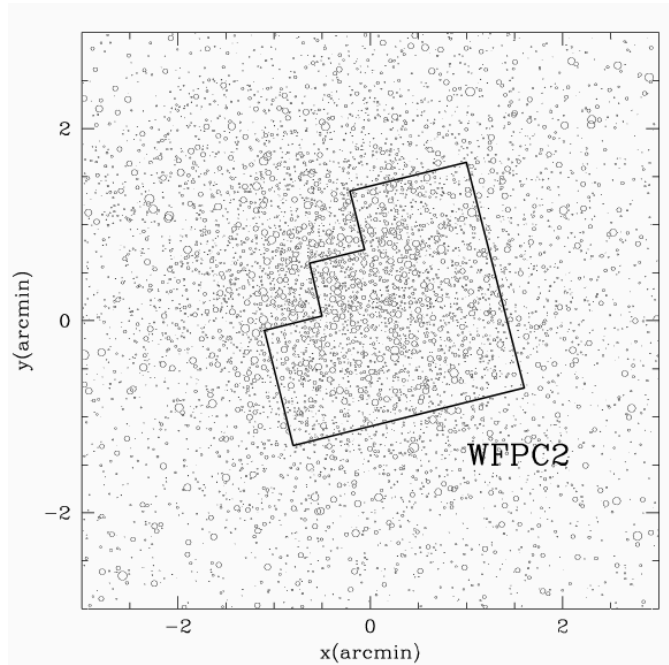}}
 \caption{\baselineskip 0.4cm %
 Plot of the location of the HST-field inside the ground-field. 
 The dimensions of the circle (representing the stars belonging to
 ground sample) depend on their magnitude. The magnitude limit of the stars 
reported in the figure is V = 21.}
 \label{fig:map} 
\end{figure}

The location of the HST field, overlapped  onto the field covered by the 
ground-based observations, is shown in Fig. \ref{fig:map}. 
A journal of observations is given in Table \ref{tab:jour}, where the 
columns contain respectively: the filter used, the number of exposures and the
exposure times in seconds for both data-sets.

\begin{table}[ht]
\begin{small}
   \caption[]{Journal of observations}
   \label{tab:jour}
$$\begin{array}{l|ccc}\cline{1-4}
{}	      & {\rm filter} & {\rm N} & {\rm t(s)}   \\ \cline{1-4}
{\rm 1.54\,Danish}  & I            & 2       & 10    \\
{}                & {}           & 1       & 30  \\
{}                & {}           & 2       & 60   \\ 
{}                & {}           & 1       & 120   \\ 
{}                & {}           & 2       & 240   \\ 
{}                & V            & 2       & 10   \\ 
{}                & {}           & 1       & 30   \\ 
{}                & {}           & 2       & 60   \\ 
{}                & {}           & 3       & 300   \\ 
{}                & B            & 2       & 60   \\ 
{}                & {}           & 1       & 180   \\ 
{}                & {}           & 3       & 600   \\ 
{\rm HST}         & F555W        & 2       & 3   	\\
{}                & {}           & 6       & 160 	\\
{}                & F814W        & 3       & 3   	 \\
{}                & {}           & 1       & 160   \\
{}                & {}           & 4       & 200    \\
{}                & {}           & 1       & 350 \\\cline{1-4}
\end{array}$$
\end{small}
\end{table}

We used the stars lying in the central region common to the
HST and the Danish field, to carefully check the independent calibrations, 
and to make the data-sets uniform.

\subsection{Data reduction}
\label{sub:reduction}

Corrections to the raw  ground-data for cosmic rays, bias, dark and 
flat-fielding were  made through the standard procedure.
On the other hand we applied DAOPHOT-II (Stetson \cite{stet87,stet92}) 
available on MIDAS for subsequent data reduction. 

The first step  of this procedure was the search of the stellar objects
on the deepest I image. 
The  detected candidates were fitted in all V, I and B frames  by
using a PSF-profile and an average instrumental magnitude was derived for 
each object and color. 

A total of 8413 objects have been detected in all the three bands 
down to $V \sim 22.5$.
Internal errors were estimated as the r.m.s. frame-to-frame scatter of
the instrumental magnitudes obtained for each star. 
The mean error in the interval $14 \leq V \leq 22$ turned out to be 
$\sigma_{\rm B}$ = 0.06 $\sigma_{\rm V}$ = 0.05 and $\sigma_{\rm I}$ = 0.04. 

Conversion from instrumental magnitudes to the Johnson standard system 
was performed by using primary calibrators (Landolt \cite{land92}), which 
covered a larger range in color than the cluster stars. 

Corrections to the raw HST-data for bias, dark and flat-fielding were
applied using the standard HST pipeline. Subsequent data reduction was made 
using MIDAS routines and the Romafot package for crowded fields (Buonanno
and Iannicola \cite{buon89}).

First, a median filter was applied to each frame in order to remove cosmic
rays from  every single frame. Then we used the deepest I image to search
for stellar objects in each WFPC2 camera.

All detected objects were fitted in all V and I frames using a PSF-profile 
modelled by a Moffat function, plus a numerical map of the residuals, to take 
into account the contribution of the stellar wings better.
The identified candidates were then measured on each individual V and I frame,
and an average instrumental magnitude was derived for each object and color. 

The instrumental magnitudes were transformed  into the standard Johnson 
system by using the Holtzman synthetic equations (Holtzman et al. 
\cite{holt95}).

\section{Color-Magnitude Diagrams}

\label{sec:cmd}

\begin{figure*}
 \includegraphics[width=10cm]{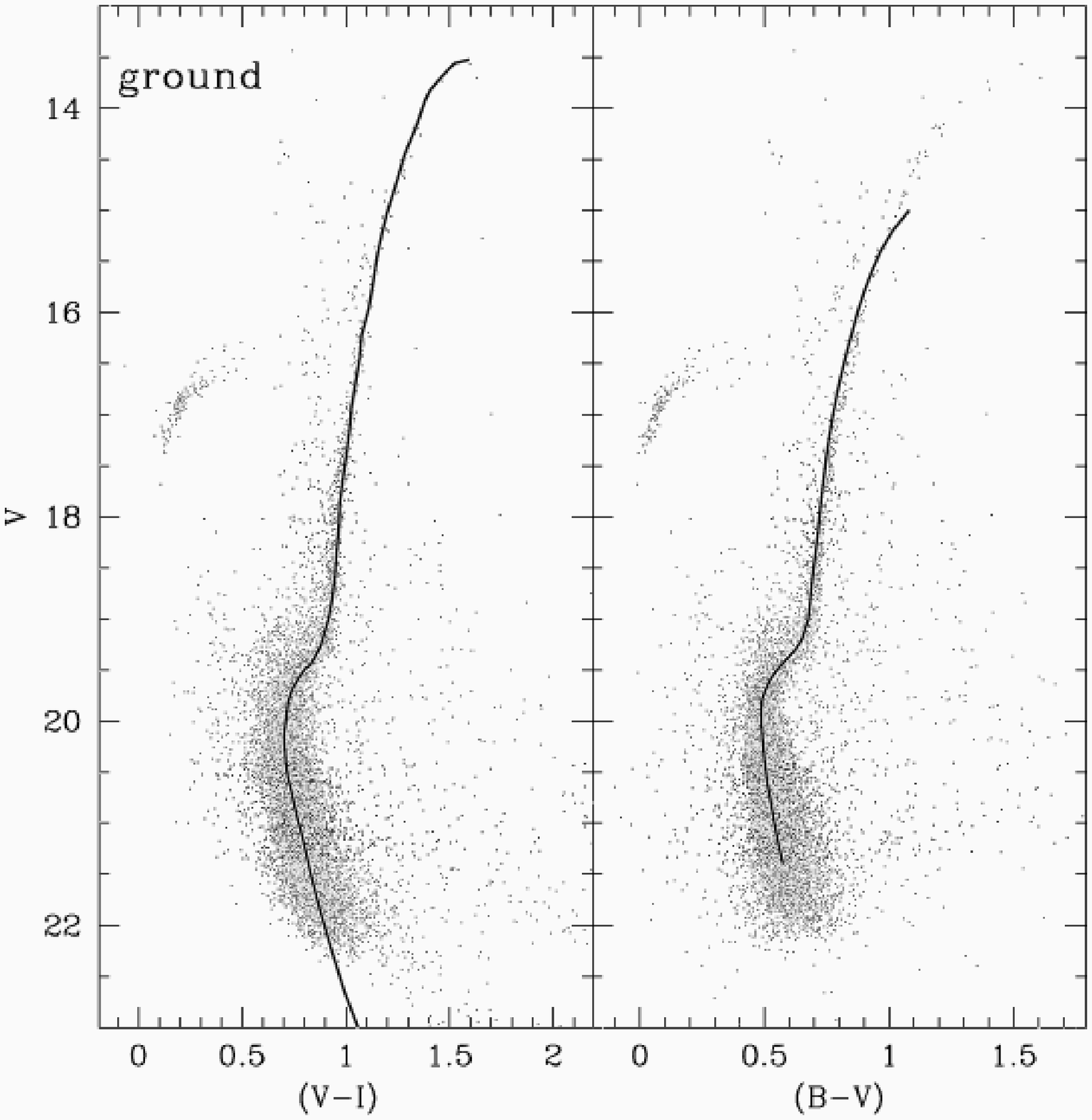}
 \includegraphics[width=10cm]{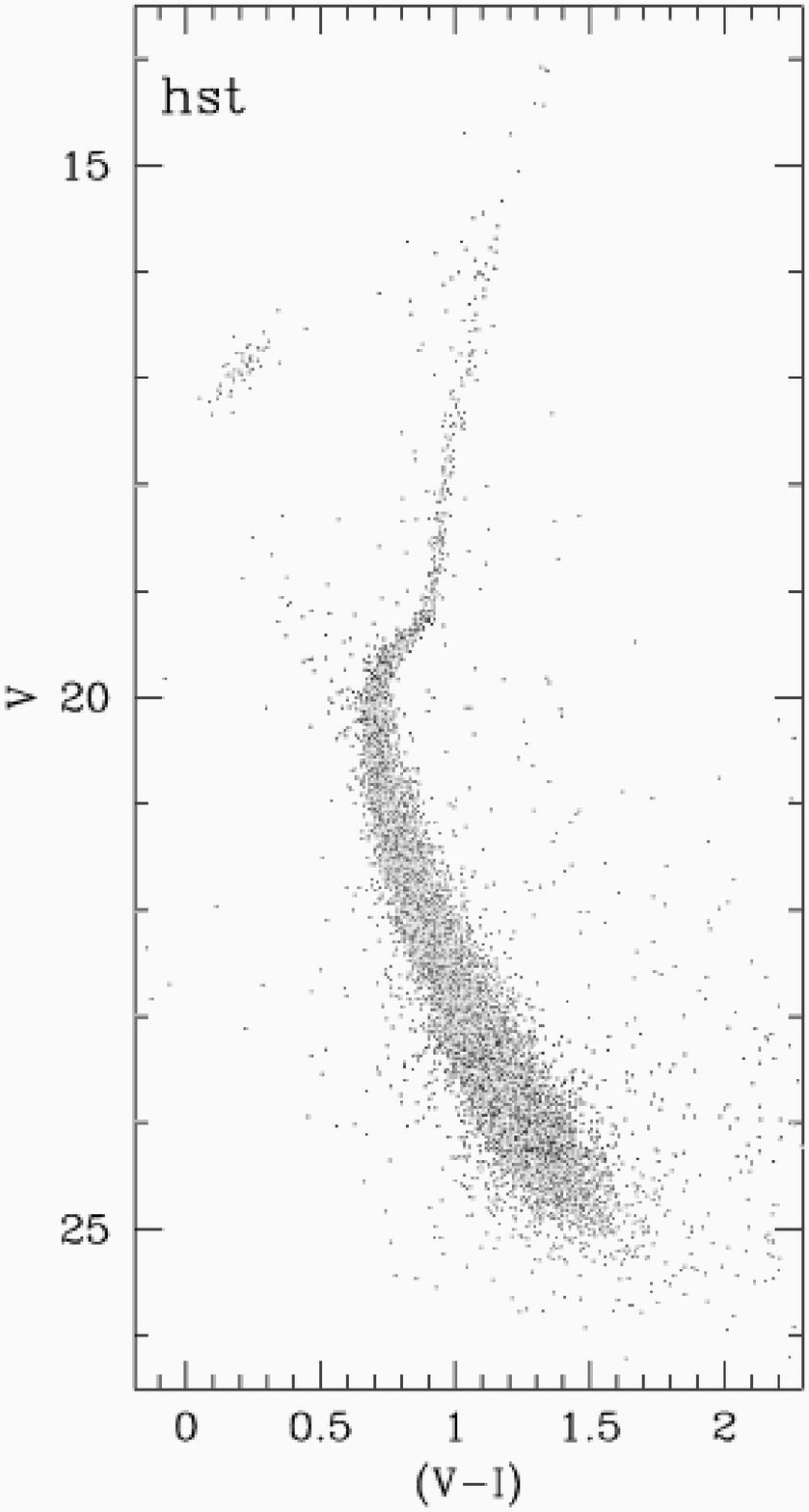}
 \caption{\label{fig:cmds} \baselineskip 0.4cm %
   Left: (V-I), (B-V) Color-Magnitude Diagrams obtained by using ground data.
  The superimposed lines are respectivaly: the ridge line obtained by using 
  HST (V-I) CMD; the fiducial (B-V) line described in Table 2 of 
  SDC.
  Right: (V-I) Color-Magnitude Diagram obtained by using HST data.
   }
\end{figure*}

Figure \ref{fig:cmds} (right panel) shows the V {\it vs} (V-I) CMD for 12012 
stars detected in the HST sample.

Figure \ref{fig:cmds} (left panels) shows:

\begin{enumerate}{}
\item the V {\it vs} (V-I) CMD for 9435 stars detected in the V and I filters 
in the ground-sample.
The overimposed curve is the ridge line obtained (i.e. see details in 
Sect. \ref{sub:3.1}) by fitting a gaussian to the distribution of (V-I) 
values in each fixed magnitude V intervals of the HST-CMD.
\item the V {\it vs} B-V CMD for 8416 stars simultaneously detected in the B 
and V filters of the ground-sample. 
The overlapped line is the fiducial sequence V {\it vs} (B-V) shown in Table 2 
of SDC.
\end{enumerate}

Because of the fainter magnitude  level reached in the central regions 
when using HST (V $\simeq$ 25.5), the number of objects present in the 
HST sample is larger than the number detected in the ground-based data, 
 even if the spatial coverage is larger for the latter (i.e. $\simeq$ 7.7 
times in area).

In the three diagrams of Fig. \ref{fig:cmds}, the richness of the samples 
allows  the identification of all the evolutionary phases. In particular, a 
well populated main sequence, subgiant and giant branches, as well as the 
less represented blue horizontal branch, blue stragglers sequence and 
asymptotic giant branch (the last actually better  seen 
in the ground based data of Fig. \ref{fig:cmds}) are clearly visible.
The overall morphology of the cluster (blue HB, steep RGB) confirms the 
metal-poor nature of NGC\,6101, as already pointed out by Zinn \& West 
(\cite{zinn84}, ZW), SDC and Rosenberg et al. (\cite{rose00}).
In the following sections, we will carry out the  analysis 
of each sequence, and we will estimate the fundamental quantities for
this cluster, namely, reddening, distance, age and metallicity.\\

\subsection{Main Sequence and Red Giant Branch}
\label{sub:3.1}

In Fig. \ref{fig:cmds} (right) we can appreciate a very narrow MS: the 
dispersion $\sigma (V-I)$, starts to grow remarkably 
($\geq$ 0.037) about one magnitude below the MSTO: V = 20.05 $\pm$ 0.05, 
(V-I) $\simeq$ 0.69. 
In particular, this value has been obtained by using the following procedure:
first  of all, we considered an interval one magnitude V wide, around the
most probable blue point of the MS in the CMD of Fig. \ref{fig:cmds} (right).
 Secondly, we divided this interval  into bins of 0.1 mags and 
computed the mean color for each bin.

The MSTO magnitude was then obtained by taking the 
central value of the bin for which we have the bluest mean color. 

In order to build a fiducial line along the RGB for ground based data, 
we assumed a Gaussian color distribution and determined the mean color and 
dispersion, $\sigma$, in each magnitude interval. In  Table \ref{tab:rgb}
the bin width, the mean, the  dispersion ($\sigma$) in color and the 
photometric errors are listed. 

\begin{table}[ht]
\begin{small}
 \caption[]{RGB fiducial line for ground data}
   \label{tab:rgb}
$$\begin{array}{l|ccc}\cline{1-4}
{\rm V}      		& {\rm < V-I >} & {\rm \sigma} & {\rm error} \\\cline{1-4}
15.50-16.00	&  1.11	& 0.02	&	0.02 \\
16.00-16.50	&  1.08 & 0.02	&	0.03	\\
16.50-17.00	&  1.05 & 0.02	&	0.03	\\
17.00-17.50	&  1.00	& 0.03	&	0.04	\\
17.50-18.00	&  0.97 & 0.03	&	0.04	\\
18.00-18.50	&  0.94 & 0.04	&	0.05	\\
18.50-19.00	&  0.93 & 0.04	&	0.05	\\
\cline{1-4}
\end{array}$$
\end{small}
\end{table}

Comparing the observed widths with the photometric errors, the former 
 appear  compatible with a null color dispersion 
along the various evolutionary phases. This fact  translates into a null 
dispersion in metallicity (Renzini \& Fusi Pecci \cite{renz88}).

\subsection{The RGB Bump}

One of the most interesting features of the RGB luminosity 
function of galactic globular clusters is the RGB bump. It appears as a peak
in the differential luminosity function and as a change in the slope of the 
cumulative luminosity function. 
From a theoretical point of view, the presence of the bump along the RGB of 
globular clusters was first predicted  by Thomas (\cite{thom67}) and Iben
(\cite{iben68}).

\begin{figure}
 \resizebox{\hsize}{!}{\includegraphics{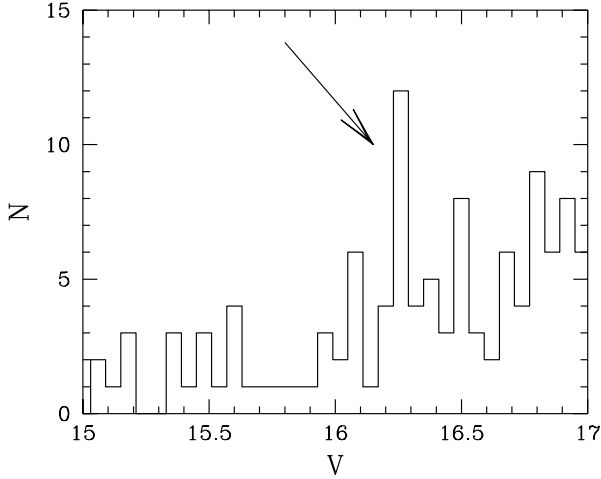}}
 \caption{\label{fig:bump_obs} \baselineskip 0.4cm 
 Differential luminosity function for the RGB in bin of 0.06 magnitude V. 
 The position of the RGB bump is marked by an arrow.}
\end{figure}

From an evolutionary point of view, the existence of the bump is due to the 
fact that during the RGB evolution, the H-burning shell crosses the chemical 
discontinuity left over by the convective envelope soon after the first 
dredge-up. 
The bump is located on the RGB at a luminosity which depends on the 
metallicity and on the age of the cluster and,  in general, is not a 
prominent feature.
In  the last few years, the RGB bump has been  a crossroad for 
several theoretical and observational investigations (see Zoccali et al. 
\cite{zocc99} and references therein). Until few years ago, the 
detection of the bump 
 was mainly hampered by the size of the available sample of RGB stars. 
This problem has been particularly relevant for the most metal-poor clusters, 
where the bump shifts toward brighter magnitudes and then less populated RGB 
regions. This is why we have tried to detect the bump in NGC\,6101 
by taking advantage of the large statistical sample of RGB stars in our 
CMDs. 
An accurate inspection of CMD of Fig. \ref{fig:cmds} (left) suggests that 
there is a bunch of RGB stars located at V $\simeq$ 16.3 and (V-I) $\simeq$ 
1.1.
In Fig. \ref{fig:bump_obs}, we have plotted the differential 
luminosity function for the RGB by adopting a bin size of 0.06 mag. 
It is worth noticing that the peak of the bump is a rather clear feature
at magnitude V=16.26 $\pm$ 0.03 mag. 

\subsection{The horizontal branch}

\begin{figure}
 \resizebox{\hsize}{!}{\includegraphics{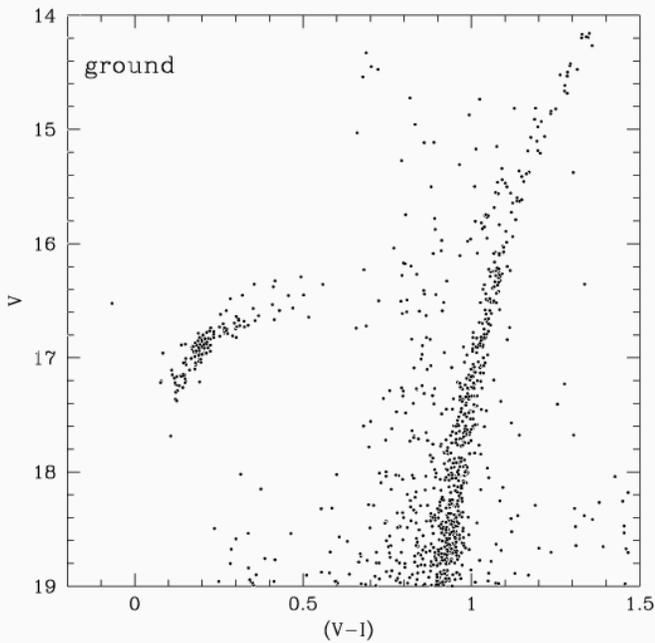}}
 \caption{\label{fig:hb} \baselineskip 0.4cm 
 Enlargement of the HB region of the (V-I) CMD of NGC\,6101
 }
\end{figure}

Starting from Alcaino (\cite{alca74}), all previous photometric studies 
have emphasized  the ``normal metal-poor'' morphology  of the horizontal
branch of NGC\,6101, showing, as expected, a blue tail.
In Fig. \ref{fig:hb} an enlargement of the HB region of the CMD of NGC\,6101 
 is shown.

 We can clearly see a populated long blue tail starting from the 
blue edge of the RR Lyrae gap (V-I)$\simeq$ 0.3 and extending for almost
0.8 magnitudes down to V $\simeq$ 17.4.
There is a  well visible gap in the distribution of the BHB stars 
starting at V $\simeq$ 17.
Following the methodology described in  Sect. \ref{sec:theo} we have 
derived for the horizontal branch $V_{\rm {HB}}$ = 16.59 $\pm$ 0.1;  this error
is mainly  caused by the difficulty to determine the luminosity level of
such a blue HB (see also  Sect. \ref{sec:theo}). 
All these values are in fairly good agreement with SDC and Rosenberg et al.
(\cite{rose00}).
   
\subsection{Blue Stragglers}

The Blue Stragglers (BSS) are the stars located 
in the region betweeen the TO and the blue tail of the HB.

Several studies (e.g. Bolte et al. \cite{bolt93} and 
Ferraro et al. \cite{ferr95}) suggested that the frequency of BSS should be
higher in the densest clusters. In the most  accepted hypothesis, BSS 
are the result of  the merging of two MS stars. In this framework,
the mass of a BSS, $m_{\rm {BSS}}$ should always be less than twice the mass of a TO
star. 

\begin{figure}
 \resizebox{\hsize}{!}{\includegraphics{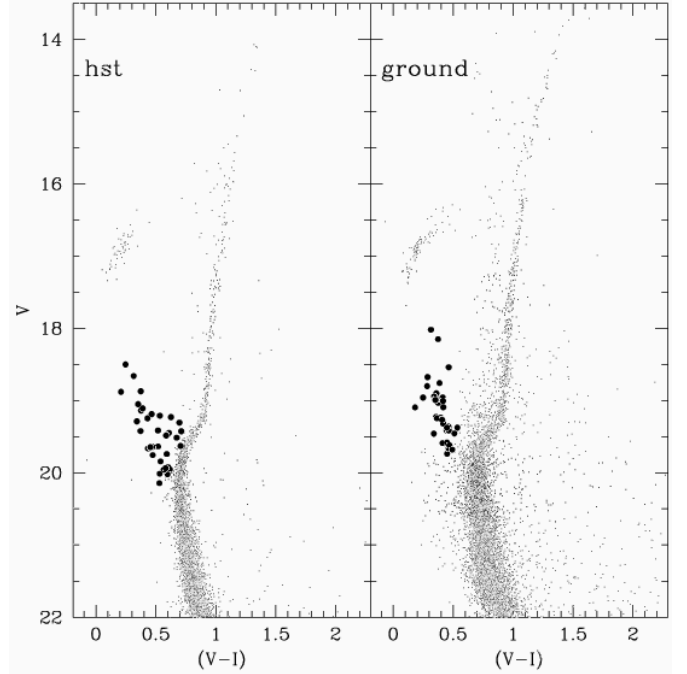}}
 \caption{\label{fig:bss} \baselineskip 0.4cm %
 Left: (V-I), Color-Magnitude Diagram obtained by using HST data.
 Black big points represent the candidate BSS identified in the HST catalog.
 Right: (V-I), Color-Magnitude Diagram obtained by using ground observations
 for stars located in the region external to HST fields.
 Black big points represent the candidate BSS identified in the ground 
 catalog in this region of the sky.
   }
\end{figure}

In Fig. \ref{fig:bss} a well defined ``sequence'' of stars 
(73 BSS) bluer and brighter than the TO stars is clearly visible. 
We have found 46 additional  BSS candidates with respect to SDC; 
this is probably due to the different spatial coverage of our frames, 
which are more central than the SDC fields. 

The candidate BSS of Fig. \ref{fig:bss} have been identified through
the following procedure:
\begin{enumerate}

\item

For the stars located in the region of the sky not  included in HST 
fields, we selected a BSS locus on the ground (V {\it vs} V-I) CMD and 
including all the objects plotted as big black points in Fig. 
\ref{fig:bss} (right).

This locus has been selected as the region between the two gaps  dividing
the region where the candidate BSS are located from the field stars region and 
from the SGB, clearly visible in the figure at (V-I $\leq$ 0.5, V $\leq$ 19.4)
and (V-I $\leq$ 0.5, V $\geq 19.7$). 

These candidates have been inspected on the original deep frames, and only the
stars detected in all deep frames and in all bands have been classified as BSS. 

\item

For the sky region covered by the HST observations, we selected the BSS locus
in the CMD obtained with HST (following the same procedure as
in 1.) and including all the objects plotted as big black points in Fig. 
\ref{fig:bss} (left).
We also identified all these objects in the ground CMD by looking at the 
position of the objects in the frames, to check for the consistency of the
sample.

In order to accept an object as a candidate BSS, we compared the 
outputs of the deep photometries coming from the 6 different HST frames in 
both filters, V and I, and for each WFPC2 camera. 
We accepted as candidates only the objects  with at least 3 measurements
in each filter and rejected the others.

The comparison among the 6 different magnitudes associated with each object 
allowed us to check the variability of the candidate BSS  as well.
We did not find any significant  variability of the candidate
BSS coming from HST observations.

\end{enumerate}  

We created a catalog of 73 BSS, 28 of which have been identified
in both ground and HST catalogs; 4 have been identified only in the HST 
catalog; 25 have been identified on ground data-sample in the region 
external to HST fields and 16 BSS have been identified in common with SDC. 
Note that, by using HST data, we found that the star classified as the BSS 
number 24 by SDC (see their Table 3), is not a real BSS but a blend
of two very close stars, identified as a single object from  the ground. 

In our analysis we also rejected 4 other objects classified as BSS from SDC
(stars 3, 4, 14 and 26 in Table 3 of SDC), because they fall on
the MS in our catalog.

\begin{figure}
 \resizebox{\hsize}{!}{\includegraphics{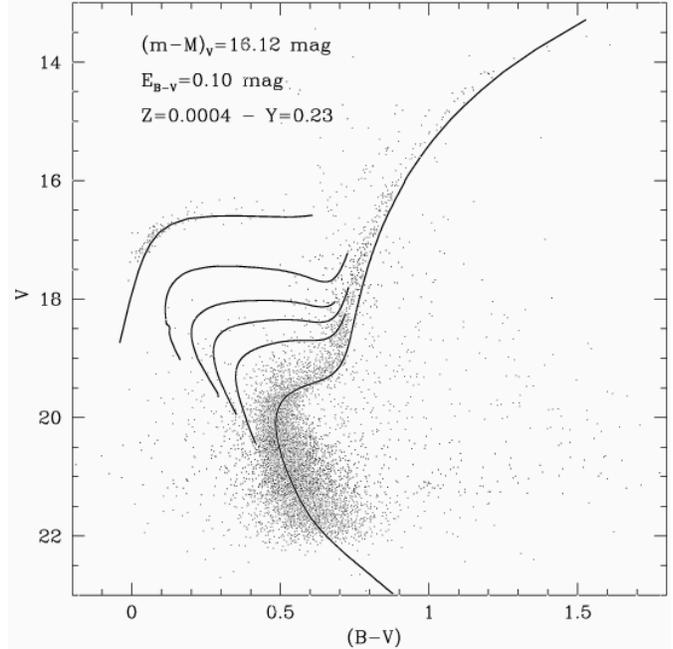}}
 \caption{\label{fig:bssteo} \baselineskip 0.4cm %
 Comparison between the (V, B-V) Color-Magnitude Diagram obtained by using 
 ground data and theoretical tracks corresponding at 1, 1.1, 1.2 and 1.4
 $M_{{\rm \odot}}$. A 13 Gyrs isochrone with a TO mass of 0.8 $M_{\rm \odot}$ is also
 shown. 
   }
\end{figure}

In the CMDs of Fig. \ref{fig:bssteo} the BSS candidates are reported together 
with: a Z = 0.0004, 13 Gyrs isochrone with a TO mass $\simeq$
0.8 $M_{\rm \odot}$ and some evolutionary tracks corresponding to 1, 1.1, 1.2 and 
1.4 $M_{{\rm \odot}}$ models, bottom to top respectively. 

It appears from the figure that the brighter BSS on the sample have a mass
$\leq$ 1.4  $M_{{\rm \odot}}$; this means that all the  BSS candidates in
NGC\,6101 are consistent with the hypothesis that $M_{\rm {BSS}} \leq$ 2$M_{\rm {TO}}$.
 
The radial distribution of these objects will be analyzed in  Sect. 
\ref{sec:radial}.

\section{Completeness and  Luminosity Function}

For HST data, incompleteness corrections were estimated as a function of the 
$I$ magnitude, by using the procedure based on selected real stars:
the real stars were added randomly to the I frames used for the initial search
of stellar peaks, paying attention to add a few  percentage
($\leq$ 10 $\%$) of the total number of stars actually present in the frames, 
to prevent an unrealistic enhancement of the image crowding.
The data reduction process described in Sect. \ref{sub:reduction} 
was then repeated from the peak detection phase to the profile fitting, 
 thus obtaining positions and instrumental magnitudes for all the objects
in the data-sample.

As a basic criterion, if a star in the output file satisfied the 
condition ($\Delta{\rm X}$, $\Delta{\rm Y} <$ 1.5 pixel, $\Delta{\rm mag} 
<$ 0.3) with respect to the input star ($N_{\rm sim}$), it was added to the number
of recovered objects ($N_{\rm rec}$). The ratio $N_{\rm rec}$/$N_{\rm sim}$ 
= $\Phi$, the completeness factor, was derived through a minimum of 10 trials 
for each bin of magnitude, then  calculating an average  within
each bin.

We built the luminosity for the three bands by dividing each CMD in 0.5 
mag-wide bins, and counting the bona fide MS stars in each bin.

\begin{figure}
 \resizebox{\hsize}{!}{\includegraphics{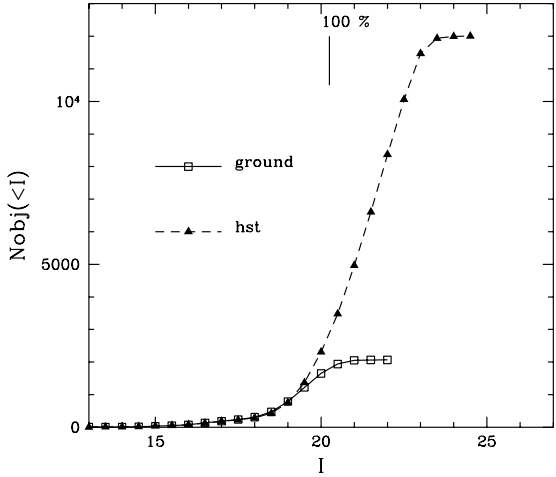}}
 \caption{\label{fig:cum} \baselineskip 0.4cm 
  Comparison between I-cumulative number counts of stars coming from 
  ground observations (for the region in common with HST observations), 
  and the I-cumulative number counts coming from HST observations.
  The percentage on the top panel is the point at which the magnitude I
  drops down 100 $\%$ (I $\simeq$ 20).}
\end{figure}

In particular, the cumulative luminosity functions in Fig. \ref{fig:cum} have
been  obtained by counting the  number of objects brighter than a fixed 
magnitude I.
This figure shows a comparison between the I-cumulative luminosity 
functions obtained by using both ground-observations (for stars located in 
the region of the sky in common with HST-observations), and HST observations.
We use the HST luminosity function to verify the 
completeness of ground data for the evolved stars (i.e. brighter than the TO 
magnitude).

The comparison between the two cumulative luminosity functions shows that 
ground-data can be considered 100 $\%$ complete down to the 
magnitude of the MSTO (I $\simeq$ 19.35), because above this 
magnitude the completeness of HST data is $\simeq$ 100 $\%$.

This is a key point in our further analysis, because it allows us to use 
ground-observations to study the radial distributions of the evolved stars 
in detail. 

\section{Radial distributions}
 
\label{sec:radial}

Because of the excellent seeing conditions of the ground observations, 
the cross-checking with HST data, and the relative looseness of NGC\,6101, 
we are confident that we sampled almost all the stars down to $I = 20$ 
inside 3 core radii ($r_{\rm c} \sim 1\arcmin$, Harris \cite{harr96}).
Thus, the comparison among radial distributions of stars belonging to the
various evolutionary phases is not affected by incompleteness.

\begin{figure}
 \resizebox{\hsize}{!}{\includegraphics{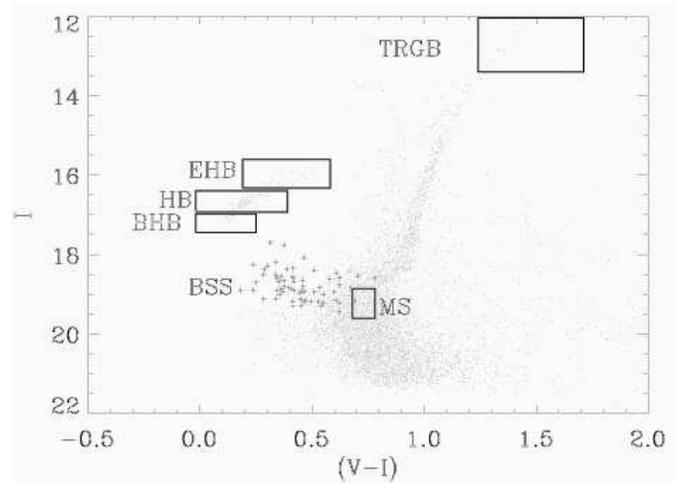}}
 \caption{\label{fig:n1}
 (I, V-I) CMD of NGC\,6101 obtained by using ground data. 
 The stellar samples adopted in order to compare their radial distributions 
 are encircled by boxes.
  }
\end{figure}

\begin{figure}
 \resizebox{\hsize}{!}{\includegraphics{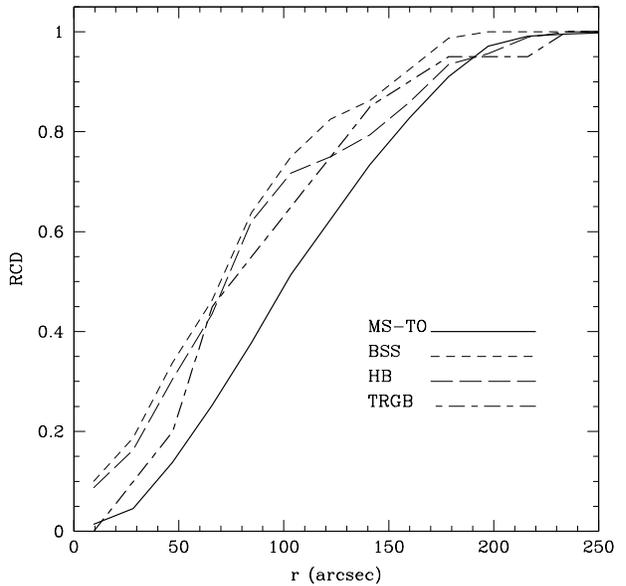}}
 \caption{\label{fig:n2}
 The radial distribution of the MS sample (continuous thick line)
 is compared with those of the RGB tip (short-long dashed),
 HB (long dashed) and BSS (short dashed).
  }
\end{figure}

Figure \ref{fig:n1} represents the stellar samples over the full area of the 
CCD ($r < 230\arcsec$), selected to compare the radial cumulative 
distributions (RCD) of the stars in different evolutionary phases.
The horizontal branch  has been subdivided in three distinct groups:
HB bluer than the gap visible at $I \sim 17$, (V-I) $\sim$ 0.1 (BHB), the core
of He-burning stars comprised between the blue gap and the RR Lyrae gap at 
(V-I) $\sim$ 0.3 (HB), and finally the supra-HB evolved stars (EHB).  
In order to check the presence of a possible over-concentration of BSS, we 
compare  - in Fig. \ref{fig:n2} - the RCDs of BSS with those of MS, 
bright tip of the red giant branch (TRGB), and HB stars.

The BSS have been selected  by following the  procedures already 
described in Sec. \ref{sec:cmd}. 
The 73 BSS (short dashed line) appear to be more concentrated than the MS stars
(continuous thick line). 

A two-sided Kolmogorov-Smirnov (K-S) test has been applied in order to 
estimate the statistical significance of the detected difference 
between couples of stellar types.
The test yields that the BSS population is more centrally
concentrated than the MS stars at the $99.9\%$ level of significance,
as well as in most galactic globulars observed at their center
(Guhathakurta et al. \cite{guha98}, Testa et al. \cite{test01}).  
This result is consistent with the hypothesis that  BSS are more massive 
binary systems (i.e. 1 -- 1.4 $M_{{\rm \odot}}$) than the bulk of the cluster 
stars. 

The bright RGB stars are more concentrated than MS-TO stars. The K-S 
probability that the RGB  stars are drawn from the same radial
distribution as the faint MS-TO stars, is less than $18\%$. 
 However, the stellar mass at the tip of the RGB is $\sim$ 
0.2 $M_{{\rm \odot}}$ lighter than TO stars, because of mass loss during the shell
H-burning phase, the TRGB stars retain  {\it memory} of the position 
inside the cluster on the basis of their  
initial mass.
Indeed, the dynamical timescale of the two-body
relaxation mechanism is much longer than the evolutionary timescale
along the RGB phase.

The HB stars (long dashed line), as defined in the box of Fig. \ref{fig:n1},
are more concentrated than MS stars, following almost the same distribution 
 of the BSS sample. 
K-S test yields a probability of $0.007 \%$ of equal radial distributions
between HB and MS, and $84 \%$ between HB and BSS. 

\begin{figure}[ht]
 \resizebox{\hsize}{!}{\includegraphics{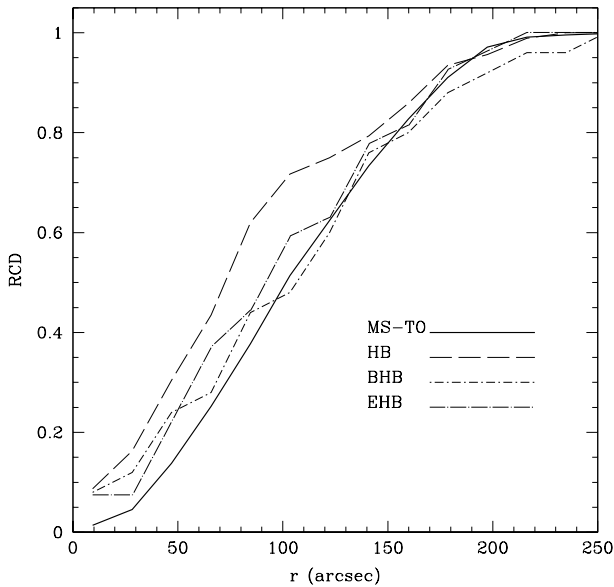}}
 \caption{\label{fig:n3}
 Comparison between cumulative radial distributions of
 HB and MS-TO stars (continuous thick line). HB stars have
 been  subdivided into three groups (see Fig. \ref{fig:n1}); 
 HB (long dashed line) is the main body of HB stars  included between the
 blue gap (V-I $\sim$ 0.2, I $\sim$ 17) and the RR Lyrae gap;
 BHB (dotted short dashed line) represents the stars bluer than the gap; 
 EHB (dotted long dashed line) indicates the HB stars evolving out of the 
 central He-burning sequence.
  }
\end{figure}    

Figure \ref{fig:n3} shows an intriguing feature of the HB's RCDs. 
The distribution of BHB stars is less concentrated than the bulk of HB objects.
K-S test gives a probability of $\sim 47 \%$ of the same radial distribution. 
This value  is even  reduced if the comparison is confined inside 
$140\arcsec$ from the cluster center. The same  result is obtained when 
the HB distribution is compared with that of EHB stars. 
EHB stars can be identified as HB stars  which leave the ZAHB phase 
starting from the very blue side of the horizontal branch (Castellani et al. 
\cite{cast91}). 
This theoretical hint is reinforced by the RCDs of both BHB and EHB. Their
radial distributions  have the same value at the $99.99\%$ level of 
significance.
This last result, together with the presence of a small gap between BHB and HB,
could be explained  by adopting the suggestion  of Buonanno et al. 
(\cite{buon97}):  stellar collisions in the dense central environment
of a cluster  might  be at the origin of the blue-sided population of the
HB. The RCDs of BHB and EHB could be explained as a further
hint of close three-body interactions in a dense environment.

Probably the ligthest stars, after a tidal-driven  extra mass-loss, have
been kicked off toward the periphery of the cluster.
However, as a matter of fact, NGC\,6101 is not a dense cluster.
Its concentration is much lower than that of 
clusters with an extended blue HB tail (Testa et al. \cite{test01}). 
Hence, why  does NGC\,6101 show the quoted peculiar radial
distribution of HB stars together with one of the highest specific 
 frequencies of the BSS, even if  it is at present a sparse cluster?

A possible answer could be found in the dynamical interactions
with the disk and/or the bulge of the Galaxy.
Takahashi \& Portegies Zwart (\cite{taka00}) have compared Fokker-Planck 
GC models, including an improved treatment of the tidal
boundary, with N-body calculations. They carried out
extensive calculations of  the dynamical evolution of clusters over a
wide range of initial conditions. They found a class of models in
which, depending on the initial parameters, the tidal dragging
of the low-mass stars drives the continuous decreasing of the total
mass and of the concentration parameter.   

Therefore,  we might now be seeing the effect of the 
``dynamical moulding'' onto a much  denser cluster.
In  this case, the BSS specific frequency, BHB and EHB radial 
distributions could be the relics of the  preceding dynamical evolution 
of NGC\,6101.

\subsection{Comparison with other clusters}

\begin{figure}
 \resizebox{\hsize}{!}{\includegraphics{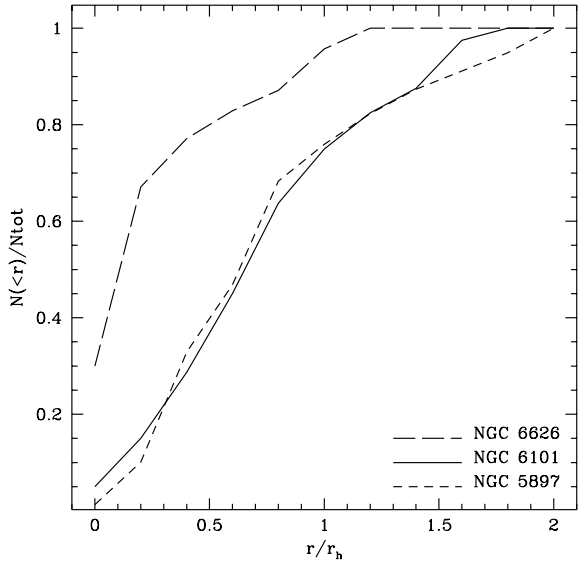}}
 \caption{\label{fig:radbss}
  BSS RCD of NGC\,6101 compared with those of NGC\,5897, and NGC\,6626.
  The RCDs are normalized to the total number of BSS and are plotted in 
  units of $r_{\rm h}$.
  }
\end{figure}

A meaningful comparison of the BSS RCDs can be only made  among
clusters in which stars are counted in sky areas defined homogeneously. 
 On this condition, we have compared the specific frequency $F^{{\rm BSS}}_{{\rm HB}}
= N_{{\rm BSS}}/N_{{\rm HB}}$ of NGC\,6101 with those of two clusters which have different
values of concentration parameter, namely NGC\,5897 ($c = 0.79$) and NGC\,6626
($c = 1.67$) (Testa et al. \cite{test01}).
In the previous formula, $N_{{\rm BSS}}$ is the number of BSS and $N_{{\rm HB}}$ is the
number of HB stars in the same area.
The ratio $0.58$ of NGC\,6101 is similar to that of NGC\,5897 ($F^{{\rm BSS}}_{{\rm HB}} =
0.48$) and more than $3$ times that of NGC\,6626 ($F^{{\rm BSS}}_{{\rm HB}} = 0.17$).
The similarity between NGC\,6101 and NGC\,5897 is noticeable also if we 
compare their RCDs.
Figure \ref{fig:radbss} reports the BSS RCDs of NGC\,6101 and those of the two
quoted comparison clusters.  
All the RCDs have been  normalized to the total number of BSS and plotted in 
units of half-mass radius ($r_{\rm h}$). As shown in the plot, the RCD of NGC\,6101 
is  very similar to that of NGC\,5897 and much less concentrated than
that of NGC\,6626.
 
These results suggest that NGC\,6101 and NGC\,5897 could have
experienced similar dynamical histories.  In fact, they do have almost 
 the same structural parameters and chemical compositions.  

The much more concentrated distribution of BSS in NGC\,6626
might suggest a different formation  mechanism with respect to
the BSS of the two less concentrated clusters, NGC\,6101
and NGC\,5897. 
According to Bailyn \& Pinsonneault (\cite{bail95}), the shape of the
LF of the BSS could be explained by invoking different formation scenarios.
 
\begin{figure}
\resizebox{\hsize}{!}{\includegraphics{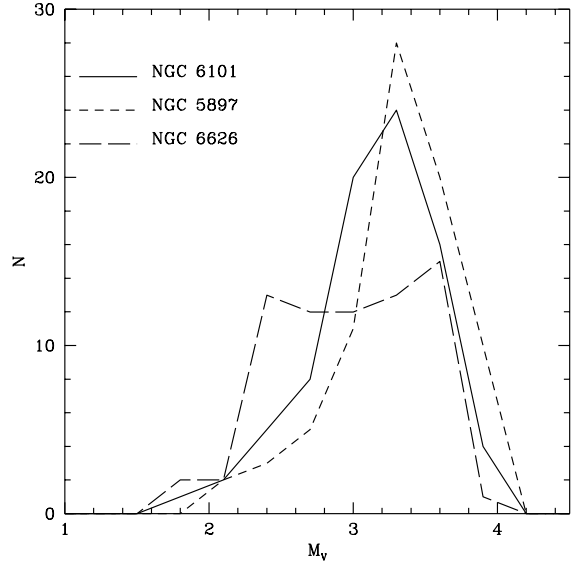}}
\caption{\label{fig:lfbss}
  BSS luminosity function in absolute $V$ magnitudes of 
  NGC\,6101 compared with those
  of NGC\,5897, and NGC\,6626.  
  }
\end{figure}    

Figure \ref{fig:lfbss} shows the luminosity functions of the
BSS in the three quoted clusters. Once again, the hypothesis of a substantial 
similarity between NGC\,6101 and NGC\,5897 is strenghtened. The shape of
their LF is comparable,  thus suggesting the same origin of their
BSS population. On the contrary, the LF of NGC\,6626 is almost
flat with a luminous tail. This favours the hypothesis of a collisional 
scenario in a dense environment as the most probable BSS formation mechanism.

\section{Comparison with theory}
\label{sec:theo}

The high quality of the photometry allows us to perform a  robust, and 
detailed quantitative comparison between the observational data for NGC\,6101
and  the theoretical predictions provided by updated stellar models.
This  combination is quite relevant in order to estimate the
age of this cluster and its distance modulus.

We have computed a set of stellar models by using the
most updated physical inputs. All the evolutionary computations
presented in this paper have been performed with the Frascati
RAphson Newton Evolutionary Code (FRANEC) (Chieffi et al. \cite{chie89}, 
Castellani et al. \cite{cast97}).
As far as the equation of state is concerned, the updated OPAL EOS 
(Rogers et al. \cite{roge96}) has been used.
In the thermodynamical regions where the OPAL EOS is not available,
it has been supplemented with the Straniero (\cite{stra88}) EOS. 
We have adopted the OPAL opacity tables (Iglesias et al. \cite{igle96}) 
combined at low temperatures with the molecular opacities provides by 
Alexander \& Ferguson (\cite{alex94}).
For a more detailed discussion on the other physical inputs adopted in
the evolutionary computations,  see the complete discussion in Cassisi et
al. (\cite{cass98,cass99}).
It is worth noticing that  the present models account for both Helium and
heavy metal atomic diffusion (Castellani et al. \cite{cast97}).
For the calibration of the superadiabatic convection we have adopted a mixing
length value of 2.0.

The color transformations and bolometric corrections used  in order to 
transform theoretical temperatures and luminosities in the BVI magnitudes are 
the ones provided by Castelli et al. (\cite{cast97a,cast97b}).

The metal abundance of NGC\,6101 has not been  well estimated yet. Until
 a few years ago, the only reliable determination was the estimate given 
by ZW ([Fe/H]=-1.81$\pm$ 0.15 dex). 
This value is in good agreement with the determination obtained by 
SDC by adopting a photometric metallicity indicator based upon the RGB
color ([Fe/H]=-1.78). 
More recently, Rutledge et al. (\cite{rutl97}), using the Ca II triplet,
have provided the metallicity of a large set of galactic
globular clusters by rescaling their evaluations to both the ZW
and Carretta \& Gratton (\cite{carr97}, CG) metallicity scale. 
For NGC\,6101, they provide a metallicity 
of [Fe/H]=-1.95 in the ZW scale and of -1.76 in the CG scale. 
 Due to the significant difference between these estimates and 
taking into account the evidence that their intrinsic uncertainty is, at 
least, of the order of 0.2 dex (see Rutledge et al. \cite{rutl97} for a 
discussion of this topic), we conservatively adopted two different metallicity
values, Z=0.0002 and Z=0.0004, for the present analysis in computing the 
stellar models. The initial He content adopted is Y=0.23.

\begin{figure}
 \resizebox{\hsize}{!}{\includegraphics{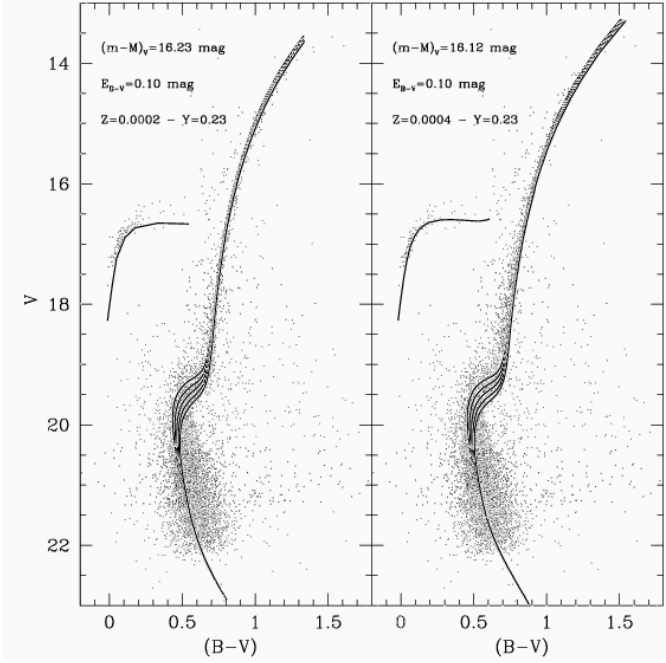}}
 \caption{\label{fig:isob} \baselineskip 0.4cm 
  Left panel: Comparison between the (V, B-V) CMD and isochrones for 
  different assumptions on the cluster age: t=10, 11, 12, 13 and 14 Gyrs, and
  for a metallicity Z=0.0002. Right panel: The same as left panel, but for 
  a metallicity Z=0.0004.}
\end{figure}

In Fig. \ref{fig:isob} (left panel), we show the comparison between the (V, 
B-V) CMD and isochrones for different assumptions  about the 
cluster age and a metallicity equal to Z=0.0002.  We have also plotted the
Zero Age Horizontal Branch locus corresponding to a RGB progenitor with mass
equal to 0.8$M_{{\rm \odot}}$. 

The cluster distance modulus has been fixed  so that the ZAHB locus 
 exactly reproduces the observed lower envelope of the stellar 
distribution along the HB turnover. 
 With this approach, we obtain a distance modulus $(m-M)_{\rm V}$=16.23 mag. 
This estimate appears in good agreement with the value given by 
SDC, $(m-M)_{\rm V}$=16.12 mag obtained through the $M_{\rm V}(RR)$ - [Fe/H]
relation provided by Lee et al. (\cite{leed90}), but it  is larger
by about 0.2 mag in comparison with the value obtained by the quoted authors
on the basis of isochrones fitting.
In Fig. \ref{fig:isob}, we have accounted for a reddening value  
E(B-V)=0.1 mag, which is in good agreement with the usually adopted estimate 
for NGC\,6101 (see SDC). From the comparison with theoretical 
isochrones, one can easily  find that the most suitable age for this 
cluster is of the order of 13 Gyrs.
In the right panel of Fig. \ref{fig:isob}, we show the same comparison, but 
with isochrones having metallicity equal to Z=0.0004. In this case, we derive 
a distance modulus equal to 16.12, in much better agreement with the estimate 
derived by SDC by using the isochrone fitting approach.   For this choice
the cluster age is of the order of 12 Gyrs.

\begin{figure}
 \resizebox{\hsize}{!}{\includegraphics{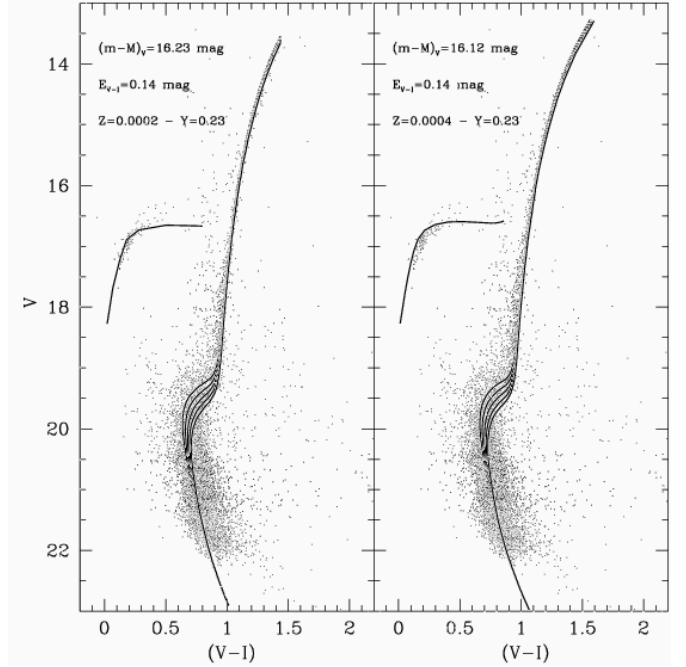}}
 \caption{\label{fig:isov} \baselineskip 0.4cm 
  Left panel: Comparison between the (V, V-I) CMD and isochrones for 
  different assumptions on the cluster age: t=10, 11, 12, 13 and 14 Gyrs, 
  and for a metallicity Z=0.0002. 
  Right panel: The same as left panel, but for a metallicity  Z=0.0004.}
\end{figure}

Figure \ref{fig:isov}, reports the same comparison for the (V, V-I) CMD.

We have used the same distance modulus and a reddening E(V-I)=0.14 mag
accounting for the relation E(V-I)=1.36 $\times{E(B-V)}$ (Taylor 
\cite{tayl86}).
It is worth noticing that, also  on this observational plane, for the
quoted choice  of distance modulus and reddening, a good agreement
exists between theoretical predictions and observational data for
a cluster age equal to $\approx$ 13 Gyrs. However, it is also  interesting
that, for the adopted distance modulus, the ZAHB locus  cannot finely 
match the lower envelope of the observed HB distribution. This  is probably
due to a drawback in the adopted color-effective temperature and bolometric 
corrections scales (see, for instance, Weiss \& Salaris \cite{weis99}). 
The same discrepancy is evident in the right panel of  Fig. 
\ref{fig:isob}, where  we show the theoretical predictions for a 
metallicity equal to Z=0.0004.

\subsection{``The Bump''}

\begin{figure}
 \resizebox{\hsize}{!}{\includegraphics{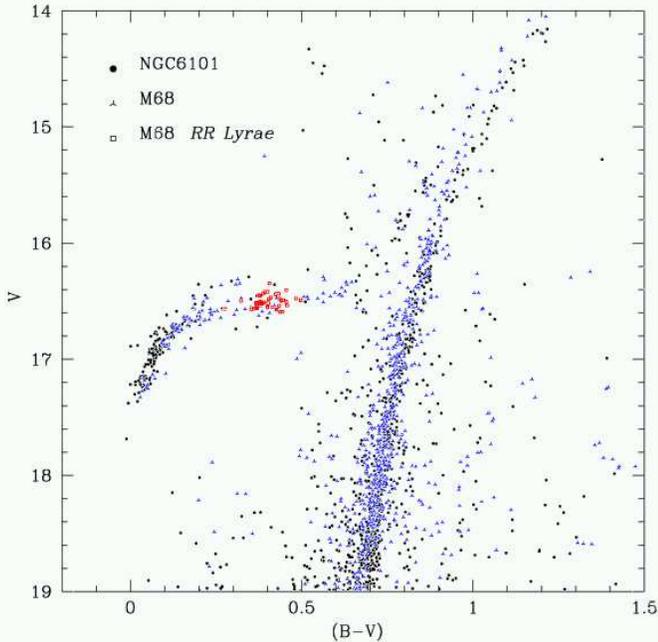}}
 \caption{\label{fig:m68} \baselineskip 0.4cm 
  Comparison between the BV-CMDs of NGC\,6101 and M68 (see text for more 
  details).}
\end{figure}

Since the work by Fusi Pecci et al. (\cite{fusi90}),  it is clear that 
a meaningful comparison between theory and observation concerning the bump 
location in the luminosity function can be performed by using the parameter 
$\Delta V_{{\rm HB}}^{\rm bump}$ - commonly defined as the difference in magnitude 
between the RGB bump and the horizontal branch stars located within the RR 
Lyrae instability strip. However, the determination of the ZAHB magnitude
$V_{{\rm ZAHB}}$ in clusters with a blue HB morphology,  as in NGC\,6101, is a 
 moot point (see Cassisi \& Salaris \cite{cass97} for more details). In 
order to overcome this problem, we have used the same approach adopted by 
Cassisi \& Salaris (\cite{cass97}).
We have chosen from the literature a cluster with metallicity
similar to NGC\,6101, with accurate photometry, a large number of RR 
Lyrae variables and a well-defined  blue HB.  The cluster M\,68 
fulfils these conditions (Walker \cite{walk94}). 
Hence, we have shifted the CMD of this cluster in color and magnitude in order
to match both the RGB and the blue HB tail of NGC\,6101. The requested shift 
in magnitude is equal to +0.87 mag (see  Fig. \ref{fig:m68}).

\begin{figure*}
 \centering
 \includegraphics[width=13cm]{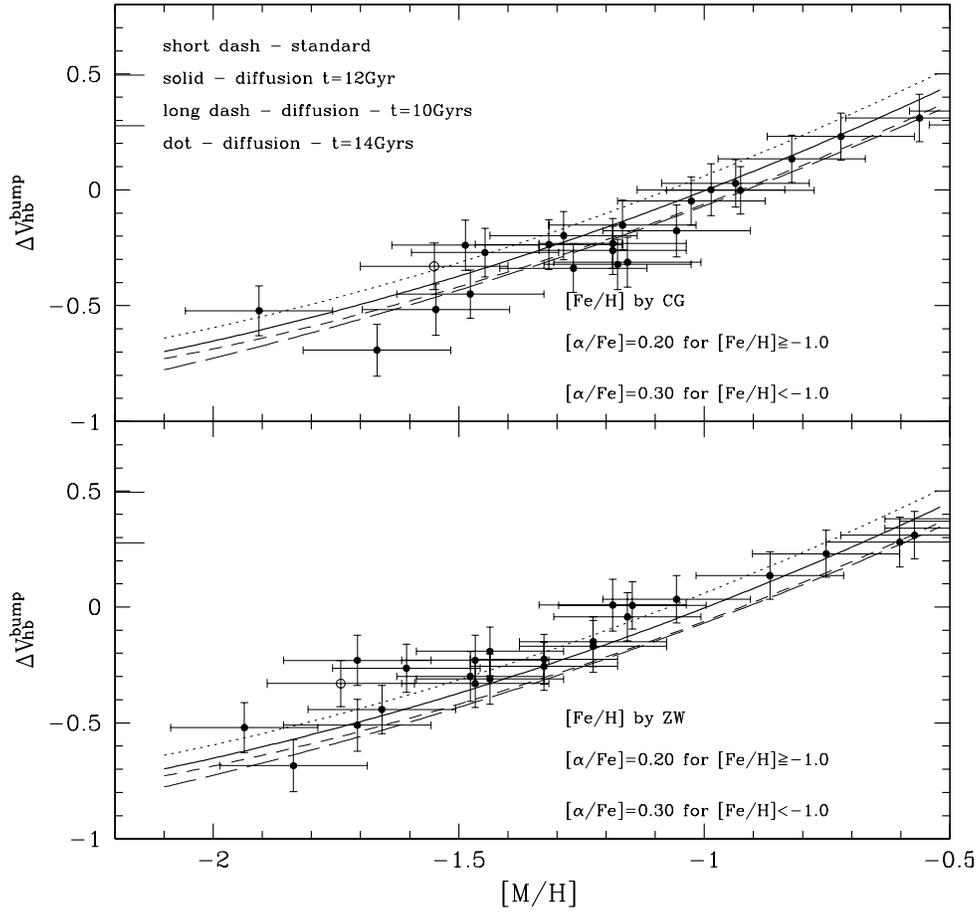}
 \caption{\label{fig:bumpteo} \baselineskip 0.4cm  Upper panel: comparison 
 between observational measurements of the parameter $\Delta V_{{\rm HB}}^{{\rm bump}}$ as 
 given by Zoccali et al. (1999), including our estimate for NGC\,6101 
 (open circle), and theoretical predictions on this quantity, by adopting the 
 metallicity scale provided by CG. 
 Lower panel: as in the upper panel but adopting the metallicity scale of 
 ZW.}
\end{figure*}

By adding this quantity to the estimated mean magnitude of the RR Lyrae in M68,
$<V_{{\rm RR}}>$ = 15.64 $\pm$0.01 mag (Walker \cite{walk94}), and by applying the 
correction between the ZAHB luminosity level and the mean magnitude of RR 
Lyrae  (Cassisi $\&$ Salaris, 1997), we have estimated a ZAHB magnitude 
within the instability strip for NGC\,6101 of $V_{{\rm ZAHB}}$ = 16.59 mag. 
 With this approach, we adopt an uncertainty on this quantity of the 
order of 0.10 mag.
Therefore, the estimated value of $\Delta V_{{\rm HB}}^{\rm bump}$ for NGC\,6101
is equal to -0.33 $\pm$ 0.10 mag. In Fig. \ref{fig:bumpteo}, we show the 
comparison between observational measurements of the parameter 
$\Delta V_{{\rm HB}}^{{\rm bump}}$ (Zoccali et al. \cite{zocc99}), including our estimate 
for NGC\,6101, and theoretical predictions of this quantity.

Accounting for the still large uncertainty affecting the globular cluster
metallicity scale, we have adopted both the ZW and the CG scales. 
In addition, we have assumed that $\alpha$-elements are enhanced in all 
clusters (see Zoccali et al. \cite{zocc99}, for more details).
In order to show the effect of the cluster age on the RGB bump magnitude, we 
have plotted relations corresponding to three different ages.  We also 
show the theoretical relation for an age equal to 12 Gyr  as provided by
canonical stellar models, i.e. not accounting for atomic diffusion.
It is worth noting that the location of NGC\,6101 appears in good 
agreement with theoretical predictions. However, the agreement is even better
 if we adopt the iron abundance suggested by ZW, which is significantly 
lower than the one provided by Rutledge et al. (\cite{rutl97}) on the CG 
scale. This result, in agreement with previous investigations performed by 
Cassisi \& Salaris (\cite{cass97}), Zoccali et al. (\cite{zocc99}) and Alves 
\& Sarajedini (\cite{alve99}), shows that the observational evidence on the 
difference in magnitude between the RGB bump and HB, also in metal-poor 
clusters, can be  well reproduced within the current theoretical 
framework.\\

\section{Summary}

We have presented  a new and complete CCD photometry for $\simeq$   
20,000 stars from the very center out to 200 $\arcsec$ inside the galactic
halo globular cluster NGC\,6101. 
The new photometry is obtained from ground based and HST data-sets.
The photometric catalogs have been used to build B {\it vs} (B-V) and V 
{\it vs} (V-I) CMDs. 
The single evolutionary phases have been analysed in detail.
We have also performed an extensive comparison of our data with the 
predictions of the stellar models.

73 BSS stars have been identified inside a radius of $\simeq$ 200 $\arcsec$.
They do not show any sign of variability. We have investigated the radial
distribution of the BSS. The BSS are more concentrated than MS-TO and
have a spatial distribution surprisingly similar to the evolved supra-HB 
stars (EHB).
Comparing the BSS content of NGC\,6101 with another similar GGC
(NGC\,5897) we found that LFs and radial distributions are
almost identical. Therefore, we argued that the mechanism at the origin
of the formation of these objects inside the two clusters should be
the same.
The large statistical sample of stars in our CMDs has allowed us to obtain
a precise detection of the RGB bump in this metal-poor cluster. 
By accounting also for the $\Delta V_{{\rm HB}}^{{\rm bump}}$ estimates 
available in the 
literature for a large database of galactic globular clusters, we have
performed a comparison between observational evidence and theoretical 
prescriptions for the luminosity difference between the RGB bump and the HB at
the level of the  instability strip of the RR Lyrae variables.
As a result, we have found a good agreement --within current 
uncertainties-- between theory and observations.

\begin{acknowledgements}
We warmly thank G. Piotto for kindly  sending us his data for NGC\,6101 and
G. Giobbi for careful reading of the manuscript. 
This work has been supported by MURST/Cofin1999 under the project: 
``Effect of the dynamics on the canonical and exotic stellar distributions 
inside galactic globular clusters''. One of us (S.C.) was supported by 
MURST/Cofin2000 under the project: ``Stellar observables of cosmological 
relevance''.

\end{acknowledgements}


\begin{thebibliography}{}
\bibitem[1974]{alca74} Alcaino, G., 1974, 1974, A\&AS 18, 9
\bibitem[1994]{alex94} Alexander, D.R., \& Ferguson, J.W., 1994, ApJ 437 879
\bibitem[1999]{alve99} Alves, D.R., \& Sarajedini, A., 
1999, ApJ 511, 225
\bibitem[1995]{bail95} Bailyn, C.D., \& Pinsonneault, M.H., 1995, ApJ 439, 705
\bibitem[1993]{bolt93} Bolte, M., Hesser, J.E., \& Stetson P.B., 1993, ApJ
408, L89
\bibitem[1989]{buon89} Buonanno, R., \& Iannicola, G., 1989, PASP 101, 294
\bibitem[1997]{buon97} Buonanno, R., Corsi, C.E., Bellazzini, M., 
Ferraro, F.R., \& Fusi Pecci, F.,  1997, AJ 113, 706
\bibitem[1997]{carr97} Carretta, E., \& Gratton, R.G., 1997, A\&A 121, 95 (CG)
\bibitem[1997]{cass97} Cassisi, S., \& Salaris, M., 1997, MNRAS 285, 593
\bibitem[1998]{cass98} Cassisi, S., Castellani, V., Degl'Innocenti, S., 
\& Weiss, A., 1998, A\&AS 129, 267 
\bibitem[1999]{cass99} Cassisi, S., Castellani, V., Degl'Innocenti, S., 
Salaris, M., \& Weiss A., 1999, A\&AS 129, 267 
\bibitem[1991]{cast91} Castellani, V., Chieffi, A., \& Pulone, L., 1991, 
ApJS 76, 911
\bibitem[1997]{cast97} Castellani, V., Ciacio, F., Degl'Innocenti, S., \& 
Fiorentini, G. 1997, A\&A 322, 801
\bibitem[1997a]{cast97a} Castelli, F., Gratton, R.G., \& Kurucz, R.L., 1997a, 
A\&A 318, 841
\bibitem[1997b]{cast97b} Castelli, F., Gratton, R.G., \& Kurucz, R.L., 1997b
A\&A, 294, 80
A\&A 324, 432
\bibitem[1989]{chie89} Chieffi, A., \& Straniero, O., 1989, ApJS 71, 47
\bibitem[1993]{djor93} Djorgovski S., 1993, in ASP Conf. Ser. 50, Structure 
and  Dynamics of Globular Clusters, eds. S. G. Djorgovski and G. Meylan, 373
\bibitem[1995]{ferr95} Ferraro, F.R., Fusi Pecci, F., \& Bellazzini, M., 1995, 
\bibitem[1990]{fusi90} Fusi Pecci, F., Ferraro, F.R., Crocker, D.A., 
Rood, R.T., \& Buonanno, R., 1990, A\&A 238, 95
\bibitem[1998]{guha98} Guhathakurta, P., Zodiac, T.W., Yanni, B., 
Schneider, D.P., \& Bahcall, J.N., 1998, AJ 116, 1757
\bibitem[1996]{harr96} Harris, W.E., 1996, ApJ 112, 148
\bibitem[1995]{holt95} Holtzman, J.A., Burrows, C.J., Casertano, S., et al., 
 1995, PASP 107, 1065
\bibitem[1968]{iben68} Iben, I. Jr., 1968, ApJ, 154, 581
\bibitem[1996]{igle96} Iglesias, C.A., \& Rogers, F.J., 
1996, ApJ 464, 943
\bibitem[1992]{land92} Landolt, A.U, 1992, ApJ, 104, 340
\bibitem[1990]{leed90} Lee, Y.-W., Demarque, P., \& Zinn, R., 1990, ApJ 350, 155
\bibitem[1986]{pryo86} Pryor, C., Smith, G.H., \& McClure, R.D., 1986, AJ 92, 
138
\bibitem[1988]{renz88} Renzini, A., \& Fusi Pecci, F., 1988, ARA\&A, 26, 199
\bibitem[1996]{roge96} Rogers, F.J., Swenson, F.J., \& Iglesias, C.A., 1996, 
ApJ 456, 902
\bibitem[2000]{rose00} Rosenberg, A., Piotto, G., Saviane, I., \& Aparicio, A.,
2000, A\&AS 144, 5
\bibitem[1997]{rutl97} Rutledge G.A., Hesser J.E., 
Stetson P.B., 1997, PASP 109, 907
\bibitem[1991]{sara91} Sarajedini, A., \& Da Costa, G.S., 1991, 
AJ 102(2), 628 (SDC)
\bibitem[1987]{stet87} Stetson, P.B., 1987, PASP 99, 191
\bibitem[1992]{stet92} Stetson, P.B., 1992, User's Manual for DAOPHOT-II
\bibitem[1988]{stra88} Straniero, O., 1988, A$\&$AS 76, 157
\bibitem[2000]{taka00} Takahashi, K., \& Portegies Zwart, S.F., 2000, 
ApJ 535, 759
\bibitem[1986]{tayl86} Taylor, B.J., 1986, ApJS 60, 577
\bibitem[2001]{test01} Testa, V., Corsi, C.E., Andreuzzi, G., Iannicola, G., 
Marconi, G., Piersimoni, A., \& Buonanno, R., 2001, AJ 121, 916
\bibitem[1967]{thom67} Thomas, H.-C., 1967, Z. Astrophys., 67, 420
\bibitem[1994]{walk94} Walker, A.R., 1994, AJ 108, 555
\bibitem[1999]{weis99} Weiss, A., \& Salaris, M., 1999, A\&A, 346, 897
\bibitem[1984]{zinn84} Zinn, R., \& West, M.J., 1984, ApJS 55, 45 (ZW)
\bibitem[1999]{zocc99} Zoccali, M., Cassisi, S., Piotto, G., Bono, G., \& 
Salaris, M., 1999, ApJ, 518, L49
\end{thebibliography}
\end{document}